\documentclass[twocolumn,prl,showpacs]{revtex4}

\usepackage{graphicx}

\begin{document}

\newcommand*{\cm}{cm$^{-1}$\/}
\newcommand{\comment}[1]{}
\newcommand*{\YBCO}{YBa$_2$Cu$_3$O$_{7-\delta}$\/}

\title{The oxygen isotope effect in the ab-plane
reflectance of underdoped YBa$_2$Cu$_3$O$_{7-\delta}$.}

\author{N.L.~Wang}
\altaffiliation[Present address: ]{Institute of Physics, Chinese
Academy of Sciences, P.O.Box 2711, Beijing 100080, P. R. China}

\author{T. Timusk}
\altaffiliation{The Canadian Institute of Advanced Research}
\email{timusk@mcmaster.ca}

\affiliation{Department of Physics and Astronomy, McMaster
University, Hamilton ON Canada, L8S 4M1}

\author{J.P. Franck}

\affiliation{Department of Physics, University of Alberta,
Edmonton, AB Canada, T6G 2J1}

\author{P. Schweiss}

\author{M. Braden }
\altaffiliation{Laboratoire L\'eon Brillouin, CE Saclay, F-91191
Gif-sur-Yvette, France} \altaffiliation[Present address: ]{II.
Phys. Inst., Univ. zu K\"oln, Zuelpicher Str. 77 D-50937 K\"oln,
Germany}
\author{A. Erb}
\altaffiliation[Present address: ]{Walther Meissner Institut,
Bayerische Akademie der Wissenschaften, D-85748 Garching,
Germany\\} \affiliation{ Forschungszentrum Karlsruhe, Institut
f\"ur Festk\"orperphysik, P.O.B. 3640, D-76021 Karlsruhe,
Germany\\}

\begin{abstract}
We have measured the effect of oxygen isotope substitution on the
ab-plane reflectance of underdoped \YBCO. The frequency shift of
the transverse optic phonons due to the substitution of $^{16}$O
by $^{18}$O yields an isotope effect of the  expected magnitude
for copper-oxygen stretching modes with $\alpha=0.5 \pm 0.1$.  The
reflectance shoulder at 400 - 500 \cm\  shows a much smaller
exponent of $\alpha=0.1 \pm 0.1$ in the normal state and
$\alpha=0.23 \pm 0.1$ in the superconducting state. These
observations suggest that the shoulder is of electronic origin and
not due to a phonon mode as has been suggested recently.
\end{abstract}


\maketitle

The absence of a substantial isotope effect on the transition
temperature of high temperature superconductors has been used to
rule out phonons as the bosons responsible for coupling between
the charge carriers\cite{franck94}. But this lack of $T_c$ shift
with isotope substitution has failed to convince everyone. Many
mechanisms in addition to the frequencies of the underlying phonon
can affect $T_c$. For example, among BCS superconductors, there
are examples where the isotope effect of $T_c$ is small and even
positive due to the Coulomb interaction parameter
$\mu^*$\cite{franck94}. whereas we know that the underlying boson
frequencies, phonons in this case, would show a square root
dependence on isotopic mass. Thus the isotope effect of $T_c$ is
of marginal value in settling the question of phonon origin of
superconductivity. A direct measurement of the isotope effect of
the boson frequency is a much more convincing test.

Recently, attention has been directed to the kink in the
dispersion of the angle-resolved photoemission spectra
(ARPES)\cite{kaminski00,valla00,bogdanov00,johnson00}. The
simplest picture of the kink is that it is the result of the
interaction of the charge carriers with a bosonic excitation. This
opens a new scattering channel at the energy required to generate
the excitation, and according to Kramers-Kronig relations applied
to the complex self energy, a kink is formed in the energy
dispersion\cite{kaminski00}. In infrared spectroscopy the kink can
be seen as an onset of absorption or a "knee" in low temperature
reflectance spectra around 500 \cm, a result of enhanced
scattering above this energy\cite{basov96c,puchkov96d}. The energy
of the knee does not vary among the various high temperature
superconductors with the notable exception of the three-layer
mercury compound where it is at a substantially higher energy, at
750 \cm\cite{mcguire00}.

The knee has been associated with the superconducting transition
(in optimally doped \YBCO\  it can only be seen below $T_c$),
although in most underdoped materials it is present in the normal
state and has therefore been related to the
pseudogap\cite{basov96c}. This has lead to the suggestion that the
bosonic excitation responsible for the knee is a possible
candidate for the pairing excitation of high temperature
superconductivity\cite{carbotte99,lanzara01}, similar to the role
played by phonons in conventional superconductivity. Recently
Lanzara \textit{et al.} have proposed that certain zone boundary
oxygen vibrations are responsible for the kink in ARPES dispersion
curves\cite{lanzara01}. A more conventional view is that the
excitation is of magnetic origin\cite{bourges98,carbotte99}.
Suggestions that it is related to the sharp 41 meV neutron
resonance have been challenged recently\cite{kee01}.

\begin{figure}
\resizebox{!}{7cm}{\includegraphics[13,200][779,800]{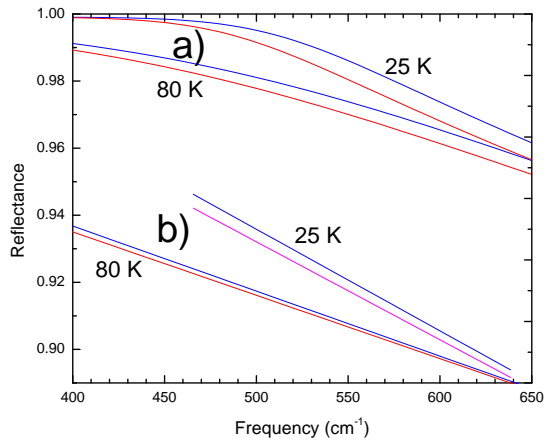}}
\caption{\label{fig1}Top curves (a), model calculation of the
reflectance change on oxygen isotope substitution at 25 and 80 K
for an copper-oxygen stretching mode phonon at a frequency of 516
\cm\ coupled to the charge carriers with $\lambda=2.2$ and
$\omega_p=1.9$ eV. Bottom curves (b) straight lines fit to
reflectance experiments on isotopically separated \YBCO. For each
temperature, two curves are shown, the top curve (dashed) is for
$^{16}$O, the bottom one (solid) for $^{18}$O.}
\end{figure}

An obvious experiment that will distinguish between these two
models of the kink excitation is the oxygen isotope effect of the
frequency of the kink. If the kink is due to phonons, its
frequency should vary as the -0.5 power-law of the reduced mass of
the mode. On the other hand, the frequency of a purely magnetic
mode should be independent of isotopic mass. Any such experiment
will be difficult because the mode giving rise to the kink is
broad. This is illustrated in Fig. 1a  where we use standard
transport theory\cite{shulga91} to calculate the conductivity and
the reflectance in a Fermi liquid system coupled to a bosonic
mode. We model the kink excitation as a copper-oxygen stretching
mode phonon of 12 \cm\ width centered at 516 cm$^{-1}$,
interacting with the charge carriers with a plasma frequency
$\omega_p=1.9$ eV and a coupling constant $\lambda=2.2$. Fig. 1a
shows that within this model, the complete replacement of $^{16}$O
by $^{18}$O in the copper oxygen planes, results in the
displacement of the knee in reflectance to lower frequency curve
by 24 \cm, or alternately, a reduction of reflectance in the
spectral region above the knee by 0.7 \%. We express a partial
isotope effect of the frequency of the mode $\nu$ in terms of the
isotope coefficient $\alpha$ defined in $d\nu/\nu = -\alpha dm/m$
where m is the reduced mass of the copper-oxygen stretching
vibration.

To our knowledge, there are no published experiments on the
isotope dependence of the frequency scale associated  with the
pseudogap state or the kink in dispersion of the  charge carriers.
There are however several reports, based on NMR and NQR
measurements, on the isotope dependence of the temperature $T^*$
associated with the onset of the pseudogap in YBa$_2$Cu$_4$O$_8$,
a material that is naturally underdoped with a $T_c=80 $ K. A zero
isotope coefficient was reported by Williams \textit{et al.}
\cite{williams98}, whereas Raffa \textit{et al.}\cite{raffa98}
found a value of $\alpha=0.061$ for $T^*$.  The same value, within
experimental error, of $\alpha=0.056$ was found for the
superconducting transition temperature $T_c$, defined as $\alpha$
in $dT_c/T_c = -\alpha dm/m$ where m is the mass of the oxygen
isotope. Giant isotope effects have been reported on the
relaxation rate of crystal field excitations\cite{temprano00}.

For the experiment we used a large single crystal of
YBa$_2$Cu$_3$O$_{7-\delta}$ which was broken in two pieces which
were parallel annealed, one in $^{16}$O and the other in $^{18}$O
to yield two underdoped samples. The weight changes of the
crystals were consistent with a near complete isotope exchange.
Neutron diffraction experiments showed that the occupancy of the
chain sites was identical for the two samples. More importantly,
the measurements also show that the z-parameter of apical oxygen
only shifts by -0.00003(4) c-axis units which is the opposite sign
from what one would expect if the shift in $T_c$ was solely due to
a difference in oxygen occupation (+0.00006)\cite{neutron-diff}.
The uncertainty in the apical oxygen position gives an uncertainty
in $T_c$ of 0.45 K. The change in shoulder frequency with Tc is
approximately 2.5 \cm/K giving an uncertainty in the shoulder
frequency due to occupancy of the chain site of only 1.1 \cm. This
is too small to compensate for a possible isotope shift expected
due to phonons (see Fig. 1) of 24 \cm.

The superconducting transition temperatures were $T_c$ = 67.6 K
for $^{16}$O and $T_c$ = 66.7 K for $^{18}$O which yields a
(partial) isotope coefficient $\alpha = 0.1$. For a pure
electron-phonon interaction in a monatomic system, $\alpha=0.5$. A
small value of $\alpha$ is typical of high temperature
superconductors\cite{franck94}. The advantage of an underdoped
sample is that it allows us to measure the kink frequency in both
the normal and the superconducting state. The samples were
twinned.

The reflectance measurements were performed on freshly cleaved
faces normal to the c-axis. To correct for irregularities of the
surface, resulting from the cleavage process, the sample was
coated with a gold layer, evaporated \textit{in
situ}\cite{homes93a}.

All the measurements were carried out in a cold-finger flow
cryostat with a minimum sample temperature of 25 K. A Bruker IFS
66 v/S spectrometer was used in the far-infrared with He-cooled
bolometer detectors below 700 \cm\ and an MCT detector up to 8000
\cm. The absolute reflectance measurements have an error of $\pm$
0.5 \%.

\begin{figure}
 \resizebox{!}{7cm}{\includegraphics[13,200][779,800]{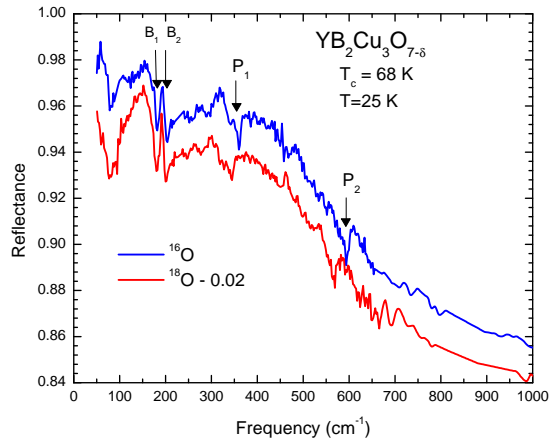}}
 \caption{\label{fig2}The reflectance of YBa$_2$Cu$_3^{16}$O$_{7-\delta}$, top
 curve, and isotopically substitute YBa$_2$Cu$_3^{18}$O$_{7-\delta}$, lower curve.
 The frequency shift of two transverse optic phonons, P$_1$ and P$_2$ can be seen
 whereas two electronic excitations, B$_1$ and B$_2$ do not shift. For clarity
 the $^{18}$O curve has been shifted down by 0.02. }
\end{figure}

Figure 2 shows the reflectance of the two crystals at 25 K. Since
the spectra are nearly identical on the scale of the figure, the
$^{18}$O curve has been displaced downward by 0.02 for clarity. We
have normalized the two spectra in the 250 to 420 \cm\ region,
below the reflectance shoulder frequency, by subtracting 0.3 \%
from the reflectance  of the $^{16}$O curve at 25 K and 0.75 \% at
80 K.

In Fig. 2 the reflectance minima marked P$_1$ (bond bending) and
P$_2$ (bond stretching) are due to transverse optic
phonons\cite{homes00}. The effect of the isotopic substitution on
their resonant frequencies can be seen clearly. From the isotopic
masses one expects a frequency ratio (FR) $\nu_{18}/\nu_{16} =
0.9428$ for pure copper-oxygen stretching vibrations. The observed
values of FR at 25 K are 0.956 and 0.953 for P$_1$ and P$_2$
respectively. The expected values from the shell model of Chaplot
\textit{et al.}\cite{chaplot95} are 0.954 and 0.9456 respectively.
The zone boundary half breathing mode is predicted to  have an
isotope ratio of 0.9437, close to the pure oxygen vibration. Note,
however, that the anomalously low frequency of the half-breathing
mode is not correctly described by the model. The good agreement
of FR with the theoretical values is consistent with complete
in-plane isotopic substitution since the isotope effect on phonon
frequencies is reduced for isotopic mixtures in a linear
fashion\cite{montgomery65}.

The modes marked B$_1$ and B$_2$ do not shift significantly with
isotopic substitution. Their frequencies and spectral weights
suggest that they may be related to the modes discussed recently
by Bernhard \textit{et al.}\cite{bernhard01}. These authors found
two modes in this spectral region with spectral weights over an
order of magnitude larger than what is expected for phonons, and
suggested that they may be of electronic origin. The absence of an
oxygen isotope shift supports this picture although, it should be
pointed out, low frequency modes generally involve heavier atoms
in the unit cell and would not be expected to exhibit a complete
oxygen isotope shift.  While the modes B$_1$ and B$_2$ have been
seen by previous workers\cite{bernhard01}, they were not seen in
ultra-pure crystals\cite{homes99} and are likely activated by
impurities.

We obtained the frequency shift of the shoulder with isotope
substitution in several ways. The most straightforward method
would be to evaluate the second derivative of the reflectance
which is closely related to $\alpha^2
F(\Omega)$\cite{allen71,farnworth74,marsiglio98,carbotte99,abanov01}
which has a peak at the frequency of the  mode responsible for the
shoulder in the conventional Fermi  liquid picture. We find this
peak at 520 \cm\ in the superconducting state and at 420 \cm\ at
80 K, in the normal state. However, as a recent study shows, even
with large crystals of Bi$_2$Sr$_2$CaCu$_2$O$_8$ which have very
flat surfaces, the error of this method is too large to resolve
the isotope effect\cite{tu01}. The source of this error is the
well known growth of error with successive differentiation. To
find a peak position in the $\alpha^2 F(\Omega)$ spectrum, we must
in effect evaluate the zero-crossing of the third derivative of
the reflectance.

To achieve the required resolution, we adopted the simpler method
of evaluating the shift along the frequency axis of the slope
portion of the reflectance in the 480 to 620 \cm\  region. As Fig.
1 shows, within the Fermi liquid model, the isotope effect gives
rise to a uniform shift of the curves along the frequency axis. We
evaluated this shift in three different ways. The simplest was to
fit a straight line to the reflectance in the slope region and to
evaluate the frequency shift of this line for the two isotopes.
This method gave $\alpha = 0.23 \pm 0.1$ in the superconducting
state at 25 K and $\alpha = 0.1 \pm 0.1$ in the normal, pseudogap
state, at 80 K. To minimize uncertainties in the values of the
absolute reflectance, all the curves were normalized in the 250 to
420 \cm\ region. In the second method, we again normalized the
curves in the flat region and and evaluated the average shift in
reflectance in the slope region and divided this by the slope of
reflectance, $dR/d\omega$ to get the frequency shift. Within an
experimental error of $\alpha$ of 0.1, the results were the same
as those obtained by the first method.

We were concerned with the influence of the optic phonons P$_1$
and P$_2$ to our analysis.  The third, more elaborate method, was
designed to subtract the optic phonons from the spectra. We
started by fitting the normalized reflectance to a series of six
Lorentzian oscillators that described the optical conductivity in
the 250 to 1000 \cm\ region. These included a Drude oscillator and
the two phonons. We then recalculated the reflectance, without the
phonons, and fitted a straight line in the slope region to the
calculated model reflectance.  The resulting isotope shift was
identical to the one found by a direct fit to the experimental
reflectance that included the phonons.

The increased isotope dependence of the shoulder in the
superconducting state may be related to the isotope effect of the
superconducting transition temperature\cite{franck94}. In our
samples $\alpha(T_c) = 0.1 \pm 0.1$. In several models of the
shoulder at $520$ \cm, its frequency in the superconducting state
is a sum of the normal state  mode frequency and the
superconducting gap frequency\cite{carbotte99,abanov01}. If there
is a proportionality between the superconducting transition
temperature and the  gap frequency, we would expect an additional
isotope effect of 0.1 in the superconducting state within the
scope of these models. This is roughly what we observed.

Another experiment that is difficult to reconcile with the phonon
model for the shoulder is the infrared reflectance of the
three-layer mercury-based superconductor which has a $T_c$ of 130
K. In this material the shoulder frequency is at 750 cm$^{-1}$,
some 45 \% higher than in materials with a maximum $T_c$ in the 94
K region\cite{mcguire00}. There are no direct measurements of the
oxygen stretch frequency in this material. However, one can
estimates this frequency from the general rule $\omega^2 \propto
a^{-3}$ where $a$ is the lattice
parameter\cite{tajima91,pintschovius89}. The lattice parameter of
the three-layer mercury compound is 3.854 at optimal doping, as
compared to 3.827 in \YBCO. The calculated phonon frequency shift,
based on the lattice parameter change, is  -0.5 \%. This is to be
compared to the observed shift of the shoulder of +45 \%.

In summary, we have measured the effect of substituting the oxygen
isotope $^{18}$O for $^{16}$O on the infrared reflectance spectrum
of underdoped \YBCO. We find that two peaks, previously assigned
to phonons, show the expected isotope shift with a coefficient
close to the theoretical value of $0.5$, whereas the reflectance
shoulder has a much smaller coefficient of $0.1 \pm 0.1$ in the
normal state which increases to $0.23 \pm 0.1$ in the
superconducting state. Thus, within experimental error, we can
rule out any simple phonon model involving copper-oxygen
stretching modes as an explanation for the reflectance shoulder in
the normal state. The small isotope shift of the shoulder suggests
that it is largely of electronic origin.

The work at McMaster University and the University of Alberta was
supported by the Canadian National Science and Engineering
Research Council. We thank J.P. Carbotte, D.N. Basov, A. Lanzara.
M.R. Norman and Z.-X. Shen for useful discussions.


\begin{thebibliography}{99}
\bibitem{neutron-diff} The structure of the two crystals was
studied on the four circle diffractometer D9 at the Institut Laue
Langevin (Grenoble), where data sets of 560 independent Bragg
reflection intensities were collected. Details of this study will
be given elsewhere.
\end{thebibliography}
\end{document}